\documentclass[aps,prd,preprint,superscriptaddress,tightenlines,nofootinbib]{revtex4}

\usepackage{graphicx}
\usepackage{dcolumn}
\usepackage{bm}

\begin{document}

\preprint{CLEO CONF 03-01}   
\preprint{CIPANP 2003}        

\title{Observation of a Narrow Resonance of Mass 2.46 GeV/c$^2$ 
       in the \\ \boldmath $D_s^{*+}\pi^0$ Final State, 
       and Confirmation of the $D_{sJ}^{*}(2317)$}

\thanks{Submitted to the 8$^{\rm th}$ Conference on the Intersections 
        of Particle and Nuclear Physics, May 2003, New York}
%
%
\author{D.~Besson}
\affiliation{University of Kansas, Lawrence, Kansas 66045}
\author{S.~Anderson}
\author{V.~V.~Frolov}
\author{D.~T.~Gong}
\author{Y.~Kubota}
\author{S.~Z.~Li}
\author{R.~Poling}
\author{A.~Smith}
\author{C.~J.~Stepaniak}
\author{J.~Urheim}
\affiliation{University of Minnesota, Minneapolis, Minnesota 55455}
% ...
\collaboration{CLEO Collaboration} 
\noaffiliation


\date{May 20, 2003}

\begin{abstract} 
  Using 13.5 fb$^{-1}$ of $e^+e^-$ annihilation data collected with 
the CLEO-II detector, we have observed a new narrow resonance in 
the $D_s^{*+}\pi^0$ final state, with a mass near 2.46 GeV/c$^2$. 
The search for such a state was motivated by the recent discovery 
by the BABAR Collaboration of a narrow state at 
2.32 GeV/c$^2$, the $D_{sJ}^*(2317)^+$, that decays to $D_s^+\pi^0$.  
Reconstructing the $D_s^+\pi^0$ and $D_s^{*+}\pi^0$ final states 
in CLEO data, we observe a peak in each of the corresponding 
reconstructed mass difference distributions, 
$\Delta M_{D_s\pi^0} = M(D_s\pi^0) - M(D_s)$ and 
$\Delta M_{D_s^*\pi^0} = M(D_s^*\pi^0) - M(D_s^*)$, 
both of them at values near 350 MeV/c$^2$.  
These peaks constitute statistically significant evidence for 
two distinct states, at 2.32 and 2.46 GeV/c$^2$, taking into account the 
background source that each state comprises for the other in light 
of the nearly identical values of $\Delta M$ observed for the two 
peaks.  We have measured the mean mass differences 
$\overline{\Delta M}_{D_s\pi^0} = 
350.4\pm 1.2\;\mbox{\rm [stat.]}\pm 1.0\;\mbox{\rm [syst.]}\;$MeV/c$^2$ 
for the $D_{sJ}^*(2317)^+$ state, and 
$\overline{\Delta M}_{D_s^*\pi^0} = 
351.6\pm 1.7\;\mbox{\rm [stat.]}\pm 1.0\;\mbox{\rm [syst.]}\;$MeV/c$^2$ 
for the new state at 2.46 GeV/c$^2$.
We have also searched, but find no evidence, for decays of 
$D_{sJ}^*(2317)$ into the alternate final states 
$D_s^{*+}\gamma$, $D_s^+\gamma$, and $D_s^+\pi^+\pi^-$.
The observations of the two states at 2.32 and 2.46 
GeV/c$^2$, in the $D_s^+\pi^0$ and $D_s^{*+}\pi^0$ decay channels 
respectively, are consistent with their possible  
interpretations as $c\overline{s}$ mesons with orbital angular 
momentum $L=1$, and spin-parity $J^P = 0^+$ and $1^+$.
\end{abstract}

\maketitle



  The BABAR Collaboration has recently reported~\cite{babar} 
evidence for a new narrow resonance with mass near 2.32\ GeV/c$^2$, 
in the $D_s^+ \pi^0$ final state.  The BABAR data is consistent with  
the identification of this state as one of the four lowest-lying 
members of the $c\overline{s}$ system with orbital angular momentum 
$L=1$, and provisionally it has been named the $D_{sJ}^*(2317)$ meson.  
A natural candidate would be the $^3P_0$ $c\overline{s}$ state 
with spin-parity $J^P = 0^+$, but other possibilities, 
including exotic states, are not ruled out.  
In this paper, we report on a search for the $D_{sJ}^*(2317)$ meson,  
as well as other, possibly related states, 
in data collected with the CLEO II detector in symmetric 
$e^+e^-$ collisions at the Cornell Electron Storage Ring, at 
center-of-mass energies $\sqrt{s} \sim 10.6$\ GeV.  

  The spectroscopy of $P$-wave $c\overline{s}$ mesons is 
summarized in Ref.\ \cite{bartelt}.  
Theoretical expectations~\cite{rgg,gi,iw,gk,dipe} 
were that: 
(1) all four states $L=1$ are massive enough that their dominant strong 
    decays would be to the isospin-conserving $DK$ and/or $D^*K$ final 
    states, 
(2) the singlet and triplet $J^P=1^+$ states could mix, and 
(3) in the heavy quark limit, the two states with $j=3/2$ would be 
    narrow while the two with $j=1/2$ would be broad, where $j$ is 
    the sum of the strange quark spin and the orbital angular momentum.
Existing experimental evidence~\cite{pdg,cleods2} for the narrow 
$D_{s1}(2536)$ and $D_{sJ}^*(2573)$ mesons which decay dominantly 
to $D^*K$ and $DK$ respectively, and the compatibility 
of the $D_{sJ}^*(2573)$ with the $J^P$ assignment as $2^+$ support 
this picture.  

  The observation by BABAR \cite{babar} of the new state is surprising  
because: (1) it is narrow (with intrinsic width $\Gamma < 10\;$MeV), 
(2) it has been observed in the isospin-violating $D_s \pi^0$ channel, 
and (3) its mass ($2316.8\pm 0.4\;\mbox{\rm [stat.]}\;$MeV/c$^2$) is 
smaller than most theoretical predictions for a $0^+$ $c\overline{s}$ 
state that could decay via this channel.  However, points 
(1) and (2) would be obvious consequences of the low mass, since the 
$D^{(*)}K$ decay modes would not be allowed kinematically.  We 
also note that at least one theoretical calculation~\cite{bh} 
had suggested that in the heavy quark limit 
the mass splittings between the $0^+$ and $0^-$ states 
of flavored mesons could be as small as 338 MeV/c$^2$, 
which is near the $D_{sJ}^*(2317)^+ - D_s^+$ mass 
splitting of $348.3\;$MeV/c$^2$ measured by BABAR.

Since the initial observation, a number of explanations have 
appeared~\cite{cahnjackson,barnescloselipkin,beverenrupp,bardeeneichtenhill,szczepaniak}. 
Cahn and Jackson~\cite{cahnjackson} 
apply non-relativistic vector and scalar exchange 
forces to the constituent quarks. Barnes, Close and 
Lipkin~\cite{barnescloselipkin} consider 
a quark model explanation unlikely and propose a $DK$ molecular state. 
Similarly Szczepaniak~\cite{szczepaniak} suggests a $D_s\pi$ atom.
Bardeen, Eichten and Hill~\cite{bardeeneichtenhill}, on the contrary, 
couple chiral perturbation theory with a quark model 
representation in heavy quark effective theory, 
building on the model described in Ref.~\cite{bh}. 
They infer that the $D_{sJ}^*(2317)$ is indeed the $0^+$ $c\overline{s}$ 
state, predict the existence of the $1^+$ partner of this state 
with a $1^+ - 0^+$ mass splitting equal to the 
$D_s^*(2112) - D_s$ ({\sl i.e.}, $1^- - 0^-$) mass splitting, and 
compute the partial widths for decays to allowed final states.
Van Beveren and Rupp~\cite{beverenrupp} also present arguments 
supporting a low mass for the $0^+$ $c\overline{s}$ state, by  
analogy with members of the light scalar meson nonet.

  The goals of the analysis presented here are to use CLEO data to 
shed additional light on the nature of the $D_{sJ}^*(2317)$,   
to provide independent evidence regarding its existence, 
and to search for decays of other new, possibly related states.    
In particular, we address the following questions.  
Are the electromagnetic decays $D_s\gamma$ or $D_s^*\gamma$ 
observable in light of the isospin suppression of the strong decay 
to $D_s\pi^0\,$?  Are other strong decays observable such as 
$D_s^*\pi^0$, or the isospin-conserving but 
Okubo-Zweig-Iizuka (OZI) suppressed~\cite{ozi} 
decay $D_s\pi^+\pi^-\,$?  If the $D_{sJ}^*(2317)$ is the 
expected $0^+$ $c\overline{s}$ state, might the remaining $1^+$ state 
also be below threshold for decay to $D^*K$, as suggested in 
Ref.~\cite{bardeeneichtenhill},  
and thus be narrow enough to be observable in its decays to 
$D_s^*\pi^0$, $D_s\gamma$ or $D_s^{*}\gamma\,$? 

This article is organized as follows.
After describing the detector and data set, 
we summarize the reconstruction of the 
$D_{sJ}^*(2317)^+ \to D_s^+\pi^0$ decay channel, including 
efforts to understand and exclude contributions from 
known background processes.  We then report on 
searches for other possible decay channels as described 
in the preceding paragraph.  We report on the appearance of 
a statistically significant signal in the $D_s^{*+}\pi^0$ 
channel, but at a location not compatible with 
$D_{sJ}^*(2317)^+$.  We describe a quantitative analysis 
of the signals in the $D_s^{+}\pi^0$ and $D_s^{*+}\pi^0$ 
channels, leading us to infer the existence of two distinct 
states.  Based on this conclusion, we discuss the 
properties of these two states, after which we summarize 
the principal results of the analysis.


  The analysis described here is based on 13.5\ fb$^{-1}$
of $e^+e^-$ collision data collected between 1990 and 1998.  
CLEO II is a general-purpose, large-solid-angle, cylindrical 
detector featuring precision charged-particle tracking and 
electromagnetic calorimetry, and is described in detail in 
Refs.\ \cite{Kubota:1992ww,Hill:1998ea}.  In its initial 
configuration, the tracking system was comprised of a six-layer 
straw tube chamber just outside of a 3.2-cm radius beryllium beam 
pipe, followed by a 10-layer hexagonal-cell drift chamber and a 
51-layer square-cell drift chamber, immersed in a 1.5-T magnetic 
field generated by a superconducting solenoid.  In 1995, the 
beam pipe and straw tubes were replaced by a 2.0-cm radius 
beam pipe plus three layers of silicon strip detectors each with 
double-sided readout, and a helium-propane gas mixture replaced 
the argon-ethane mixture previously used in the main drift chamber.  

  Beyond the tracking system but within the solenoid were also 
located a 5-cm thick plastic scintillation counter system for 
time-of-flight measurement and triggering, as well as a barrel calorimeter 
consisting of 6144 tapered CsI(Tl) crystals 30-cm in length and arrayed 
in a projective geometry with their long axis oriented radially with 
respect to the $e^+e^-$ interaction point.  An additional 1656 
crystals were deployed in two end caps to complete the solid angle 
coverage.  The excellent energy and angular resolution of the 
calorimeter is critical for the reconstruction of 
$\pi^0\to\gamma\gamma$ decays as well as single low-energy photons 
such as those emitted in the $D_s^{*+}\to D_s \gamma$ transition.

  The search for the $D^{*}_{sJ}(2317)$ was carried out by 
reconstructing the $D_s^+ \pi^0$ state, using the $D_s^+ \to 
\phi\pi^+$ channel with $\phi\to K^+K^-$.  
Charge conjugation is implied throughout this report.
Pairs of oppositely-charged 
tracks were considered as candidates for the decay products of 
the $\phi$ if the specific ionization ($dE/dx$) is measured 
in the main drift chamber to be within 2.5 standard deviations 
of expectations for kaons, and if the invariant mass of 
the $K^+K^-$ system was within $\pm 10\;$MeV/c$^2$ of the $\phi$ mass.
A third track with $dE/dx$ consistent with the expectation for a pion 
was combined with the $K^+K^-$ system to form a $D_s^+$ candidate.
The observed $D_s^+$ peak has a Gaussian width ($\sigma$) 
of $6.4\pm 0.2\;$MeV/c$^2$ in our data, consistent with CLEO Monte Carlo 
simulations of $D_s$ 
production and decay plus a GEANT-3~\cite{geant} 
based simulation of particle propagation and detector response.

%

Pairs of clusters of energy deposition of greater than 100 MeV apiece, 
of which one cluster must lie  
in the central region of the calorimeter ($|\cos \theta| < 0.71$, 
where $\theta$ is measured with respect to the beam axis) were 
selected as the candidates for photons from $\pi^0$ decay if they 
satisfied $122 < M(\gamma\gamma) < 148\;$MeV/c$^2$.  
The $M(\gamma\gamma)$ 
distribution for photon-pairs accompanying a $D_s^+$ candidate 
with $M(KK\pi)$ between 1.955 and 1.979 GeV/c$^2$ 
has a width of $5.8\pm 0.4\;$MeV/c$^2$ in our data, consistent with 
expectations from the Monte Carlo simulations.

\begin{figure}
  \includegraphics*[width=5.in]{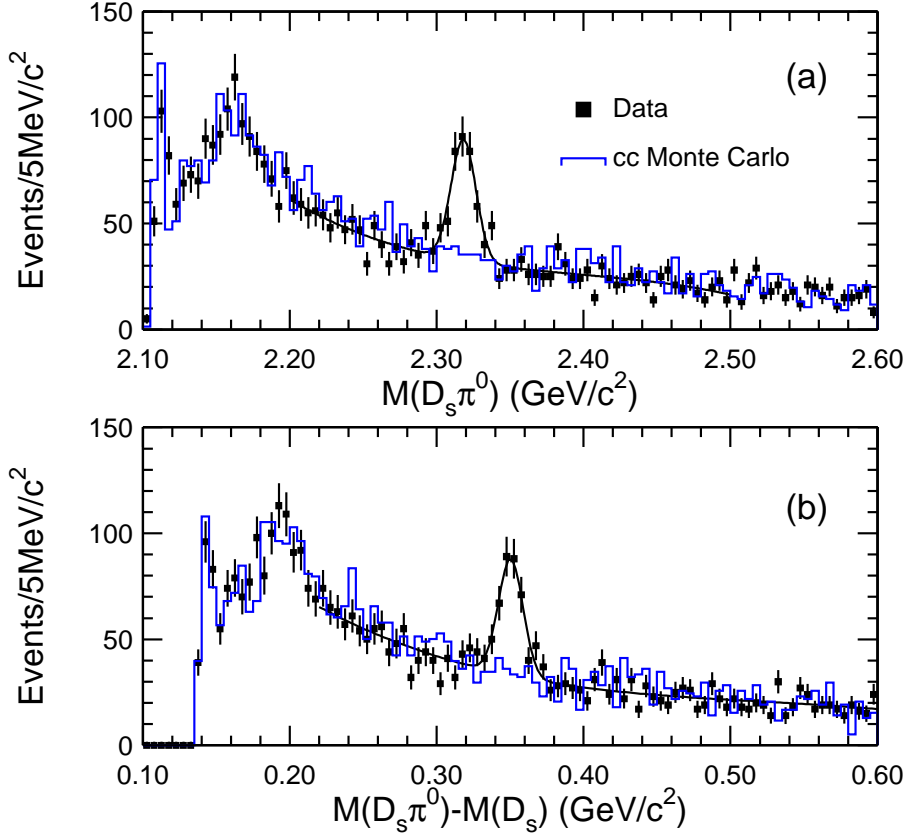}
  \caption{The $M(KK\pi\pi^0)$ mass and $M(KK\pi\pi^0) - M(KK\pi)$ 
           mass difference
           for events satisfying cuts on $M(KK\pi)$ consistent with the 
           $D_s$ and $M(\gamma\gamma)$ consistent with the $\pi^0$, 
           as described in the text.  The points represent the CLEO data, 
           while the solid histogram is the predicted spectrum from the 
           Monte Carlo simulation of $e^+e^-\to c\overline{c}$ events.  
           The predicted spectrum is normalized to the data in the 
           region between 2.50 and 2.60 GeV/c$^2$.  
           The overlaid curve 
           represents the results from a fit of the data to a Gaussian 
           signal function plus a polynomial background function.
           }
  \label{fig:manddeltam}
\end{figure}

To suppress combinatoric backgrounds, we further required that the 
momentum of the $D_s^+\pi^0$ candidate be greater than $3.5\;$GeV/c.  
We also required that the helicity angle of the $\phi\to K^+K^-$ decay 
satisfy the requirement $|\cos{\theta_h}| > 0.3$, where $\theta_h$ is 
the angle between the $K^+$ momentum vector, as measured 
in the $\phi$ rest frame,  and the $\phi$ momentum vector, 
as measured in the $D_s$ rest frame.  
The expected distribution from real $\phi$ decays varies as 
$\cos^2{\theta_h}$, whereas combinatoric backgrounds tend to be flat.
For $D_s\pi^0$ combinations satisfying these requirements, 
we plot the mass $M(KK\pi\pi^0)$ and the mass difference $\Delta M = 
M(KK\pi\pi^0) - M(KK\pi)$ in Fig.\ \ref{fig:manddeltam}. 
To improve the experimental resolution on $M(KK\pi\pi^0)$, 
the true $D_s$ mass has been used to determine the 
energy of the $KK\pi$ system from its measured 
momentum in Fig.~\ref{fig:manddeltam}(a);
this substitution is not done for $\Delta M$ 
in Fig.~\ref{fig:manddeltam}(b), or for the 
calculation of other mass differences entering this 
analysis.

The narrow peaks in Fig.\ \ref{fig:manddeltam} at mass near 2.32 GeV/c$^2$ 
and $\Delta M$ near 350 MeV/c$^2$ are in qualitative 
agreement with the BABAR observation.  We note that there are no 
peaks in this region when $KK\pi$ combinations with $M(KK\pi)$ lying 
in $D_s$ side bands are combined with a $\pi^0$.  The other features in 
the spectra shown in Fig.\ \ref{fig:manddeltam}, 
namely the sharp signal from $D_s^{*+}\to D_s^+\pi^0$~\cite{cleodspi0}
near the kinematic threshold and the broad enhancement above this 
(due primarily to $D_s^{*+}\to D_s^+\gamma$ plus a random photon, as 
well as a small contribution from $D^+ \to K^-\pi^+\pi^+\pi^0$ with a 
mass mis-assignment), 
correspond well with the features present in the BABAR data.  In 
addition, Monte Carlo simulations of inclusive charmed hadron 
production via $e^+e^-\to c\overline{c}$ give $M(D_s^+\pi^0)$ 
and $\Delta M$ 
spectra that reproduce the features observed in the data, except for 
the peak at 2.32 GeV/c$^2$ and 350 MeV/c$^2$.  This is also illustrated in 
Fig.\ \ref{fig:manddeltam}, where normalizations of each of 
the $c\overline{c}$ Monte Carlo spectra are fixed so as to match the 
last 20 bins of the corresponding data spectrum.

We have investigated mechanisms by which a peak at 2.32 GeV/c$^2$ 
could be generated from decays involving known particles, 
either through the addition, omission or substitution 
of a pion or photon, or through the mis-assignment of 
particle masses to the observed charged particles.  In no 
cases were narrow enhancements in the $M(D_s^+\pi^0)$ spectrum 
near 2.32 GeV/c$^2$ observed.  We will discuss the issue of 
feed down from a possible new resonance at 2.46 GeV/c$^2$ when we 
describe our studies of the $D_s^{*+}\pi^0$ final state.

  From a fit to the $\Delta M$ distribution
with a Gaussian signal shape and 3rd order polynomial background 
function, we obtain a yield of $231^{+31}_{-29}$ events 
in the peak near 350 MeV/c$^2$.  From Monte Carlo simulations, 
the detection efficiency associated with the reconstruction of the 
full $D_{sJ}^*(2317)^+\to D_s^+\pi^0$, $D_s^+\to\phi\pi^+$, 
$\phi\to K^+K^-$ decay chain 
is $(13.1\pm 0.7)\,\%$ for the portion of the 
$D_{sJ}^*(2317)^+$ momentum spectrum above $3.5\;$GeV/c, 
where this efficiency does not include the $D_s$ and $\phi$ 
decay branching fractions.

  The fit returns a mean and Gaussian width for the $D_{sJ}^*(2317)$ 
peak of $\overline{\Delta M}  =  350.3\pm 1.0\;$ MeV/c$^2$  and 
$\sigma = 8.4^{+1.4}_{-1.2}\;$MeV/c$^2$, respectively, where the 
errors are due to statistics only.
The expected mass resolution is $6.4\pm 0.4\;$MeV/c$^2$, somewhat 
smaller than the width returned from the fit to the CLEO data. 
Further discussions of the broader-than-expected 
width, as well as of systematic 
errors in the measurements of the mass and width of the 
$D_{sJ}^*(2317)$ appear later in this article.  However, 
the mean mass difference and width of the peak are 
consistent with the corresponding BABAR values~\cite{babar}. 
Thus, we confirm the existence of a peak in the $D_s\pi^0$ 
mass spectrum that cannot be explained as reflections from 
decays of known particles.


  The conclusion that the $D_{sJ}^*(2317)$ is a new narrow 
resonance decaying to $D_s\pi^0$ leads to two questions:  
(1) are there other observable decay modes, and (2) might  
additional new $c\overline{s}$ resonances  
also exist in which normally suppressed decay modes such as 
$D_s^{(*)}\pi^0$ are dominant?  To answer these questions we have 
searched in the channels $D_s\gamma$, $D_s^*\gamma$, 
$D_s^*\pi^0$ and $D_s\pi^+\pi^-$.  

  If the $D_{sJ}^*(2317)$ is a $0^+$ $L=1$ $c\overline{s}$ 
meson as has been suggested~\cite{bardeeneichtenhill}, it could 
decay via $S$- or $D$-wave to $D_s^*\gamma$,  
but would not be able to decay to $D_s\gamma$ due to parity and 
angular momentum conservation.  Consequently,  observation of one 
or the other of these channels would be interesting.  
On the other hand, if neither channel is seen, this would 
not be too surprising since these are electromagnetic 
decays, and the $D_s\pi^0$ decay, while isospin-violating, 
is not as severely phase-space suppressed as in the case of 
the corresponding decay of the $D_s^*$ where the 
electromagnetic decay dominates.  The BABAR data show no 
evidence for either channel, however no upper limits were 
reported on the branching ratios for these channels.  

With regard to strong decays, 
the $D_s\pi^+\pi^-$ final state is kinematically allowed and  
isospin-conserving, but would be suppressed by the 
OZI rule.  This is in contrast to the $D_s\pi^0$ 
channel for which one mechanism would be decay to a $D_s$ plus a 
virtual $\eta$, with production of the $\pi^0$ via $\eta$-$\pi^0$ 
mixing~\cite{cho}.  Observation of the $D_s\pi^+\pi^-$ channel 
would be strong evidence against the interpretation of the 
$D_{sJ}^*(2317)$ as a $0^+$ meson.  

  Finally, it is possible that the remaining $L=1$ 
$c\overline{s}$ state with $J^P = 1^+$ could also be light 
enough that decays to $D^* K$ would be kinematically forbidden.
In this case, the strong isospin-violating 
decay of this $1^+$ state to $D_s^*\pi^0$ 
could occur via $S$-wave (the electromagnetic decays to 
$D_s\gamma$ or $D_s^{*} \gamma$ 
would also be possible), and thus a peak in the 
$\Delta M_{D_s^*\pi^0} = M(D_s\gamma\pi^0) - M(D_s\gamma)$ 
spectrum would be a compelling signature of such a state.

\begin{figure}
  \includegraphics*[width=5.0in]{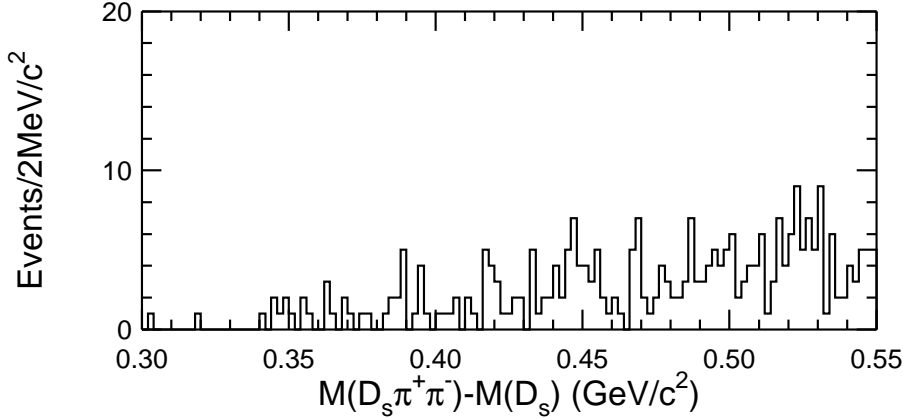}
  \caption{The mass difference $M(D_s\pi\pi) - M(D_s)$
           for $D_{sJ}^*(2317)^+ \to D_s^+\pi^+\pi^-$ candidates, 
           as described in the text.
           }
  \label{fig:deltampipi}
\end{figure}

  To look for these channels we select events containing 
$D_s^+\to\phi\pi^+$ candidates as in the $D_s\pi^0$ analysis.
For the $D_s\pi^+\pi^-$ channel, we combine the $D_s$ candidates 
with two oppositely charged tracks, and plot the mass difference 
$\Delta M_{D_s\pi\pi} = M(D_s\pi\pi) - M(D_s)$, where $M(D_s) 
= M(KK\pi)$, as shown in Fig.~\ref{fig:deltampipi}.
No signal is evident.

\begin{figure}
  \includegraphics*[width=5.0in]{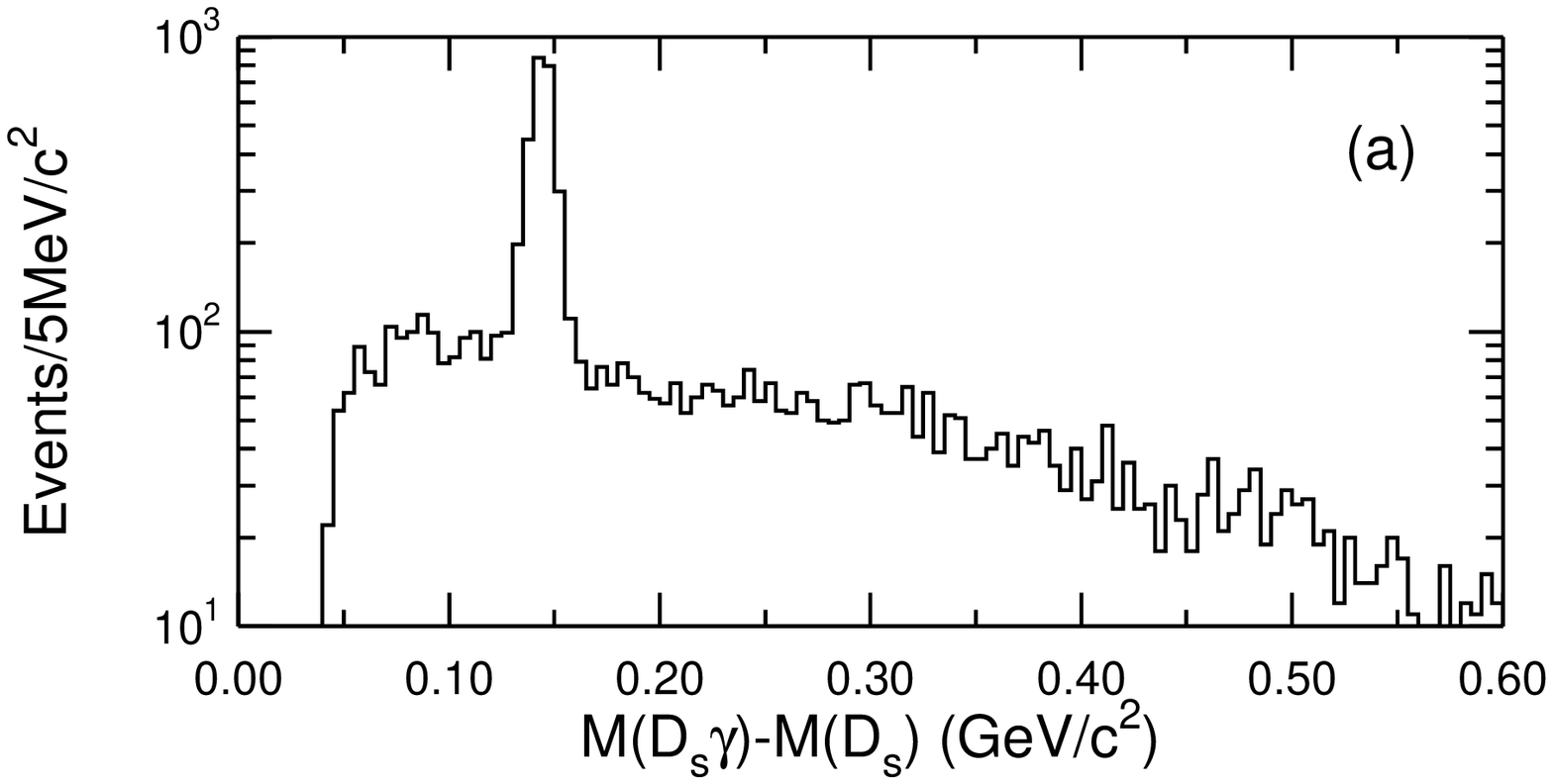}
  \includegraphics*[width=5.0in]{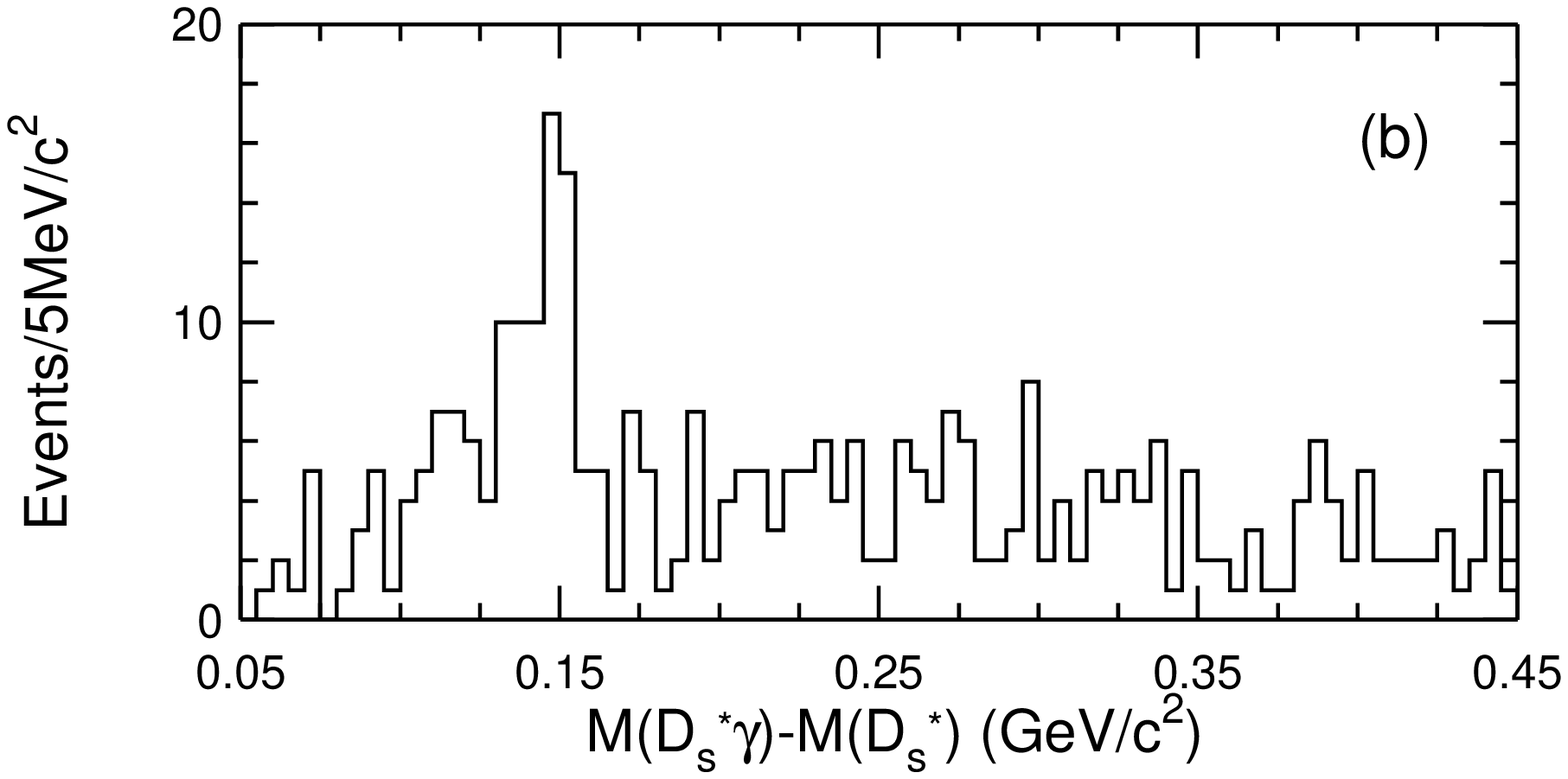}
  \caption{(a) Spectrum of the mass difference $M(KK\pi\gamma) - M(KK\pi)$, 
           plotted on a logarithmic scale.  The peak is due to 
           the transition $D_s^{*+} \to D_s^+ \gamma$.   
           (b) Spectrum of the $M(D_s^* \gamma) - M(D_s^*)$ mass 
           difference for $D_s^*\gamma$ candidates.
           }
  \label{fig:deltamgamma}
\end{figure}

To search for states decaying to $D_s^+ \gamma$, as well as 
to select $D_s^{*+}$ candidates for use in other searches, 
we have formed $D_s^+\gamma$ combinations by selecting photons 
of energy greater than 150 MeV. 
We ignore photons that can be paired with another photon 
such that $M(\gamma\gamma)$ is consistent with $\pi^0$ decay.
The inclusive $\Delta M_{D_s\gamma} = M(KK\pi\gamma) - M(KK\pi)$ 
spectrum for this sample is plotted in Fig.\ \ref{fig:deltamgamma}(a), 
illustrating that a large $D_s^*$ sample can be obtained.  For 
decay modes with a $D_s^{*}$ in the final state, 
We select $D_s\gamma$ combinations where $M(D_s\gamma)$ is 
reconstructed to be between 2.090 and 2.130 GeV/c$^2$.

Also visible in Fig.~\ref{fig:deltamgamma}(a),  are 
regions of the $\Delta M_{D_s\gamma}$ spectrum 
where decays of the $D_{sJ}^*(2317)$ (or of 
a possible higher mass state) into $D_s\gamma$ would appear.
There is no evidence for a signal corresponding to 
a $M(D_s\gamma)$ in the vicinity of 2.32 GeV/c$^2$, or elsewhere 
in the plotted region.  

The same conclusion holds for
the $D_s^*\gamma$ final state, shown in Fig.~\ref{fig:deltamgamma}(b), 
where we combine selected $D_s^*$ candidates 
with photons of energy above 200 MeV. 
The peak in the $\Delta M_{D_s^*\gamma}$ spectrum in 
Fig.~\ref{fig:deltamgamma}(b) near 150 MeV is due to real 
$D_s^{*+} \to D_s^+\gamma$ decays in which a random photon 
has been combined with the $D_s^+$ candidate to form the 
$D_s^*$ candidate, and the actual photon from this transition 
is combined with this system to form the $D_{sJ}^*$ candidate.  
There is no sign of any additional structure in this spectrum.

\begin{figure}
  \includegraphics*[width=5.0in]{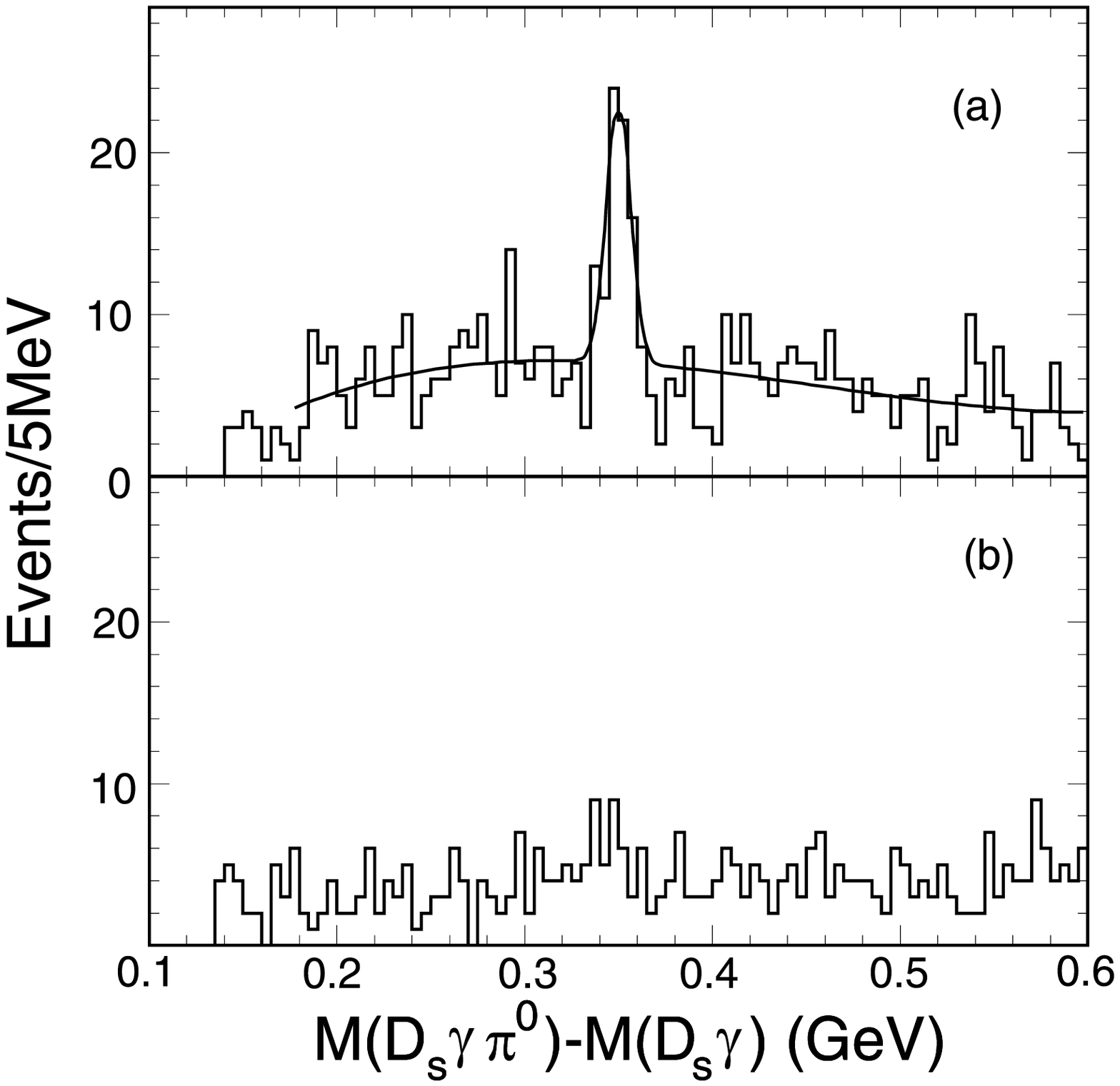}
  \caption{(a) The mass difference spectrum 
           $\Delta M_{D_s^*\pi^0} = M(D_s\gamma\pi^0) - M(D_s\gamma)$
           for combinations where the $D_s\gamma$ system is consistent
           with $D_s^*$ decay, 
           as described in the text.  (b) The corresponding spectrum 
           where $D_s\gamma$ combinations are selected from the 
           $D_s^*$ side band region.
           }
  \label{fig:deltamstar}
\end{figure}

We have also searched in the $D_s^{*+}\pi^0$ channel for $D_{sJ}^*$ states. 
This analysis was carried out with different selection criteria than
the channels described above.
Fig.~\ref{fig:deltamstar}(a) shows the mass difference plot for events 
with candidate $D_s^+\to \phi\pi^+$, $D_s^{*+}\to \gamma D_s^+$ and 
di-photon combinations consistent with being $\pi^0$ candidates. 
Other requirements are similar to those described in preceding 
sections of this paper, except that all photon 
candidates are required to be in the central region of the 
calorimeter, and we do not veto extra photons consistent
with $\pi^0$ decay to maintain efficiency.
As before, the $D_s^*\pi^0$ candidates are required 
to have momenta above 3.5 GeV/c.  A signal is 
apparent at a $\Delta M$ of $350.6 \pm 1.2\;$ MeV/c$^2$, with a width of 
$6.1 \pm 1.0\;$MeV/c$^2$ consistent with our mass resolution 
of $6.6\pm 0.5\;$MeV/c$^2$. 
A fit to a Gaussian signal plus polynomial background yields 
$53.3 \pm 9.7$ signal events.  If the $D_{sJ}^*(2317)^+$ 
were to decay to the $D_s^{*+}\pi^0$ final state, a peak would 
be expected at a $\Delta M$ of $\sim 205\;$MeV/c$^2$.  The existence 
of a peak at 351 MeV/c$^2$ leads us to investigate the possibility of a 
second narrow resonance with a mass near 2463 MeV/c$^2$ that decays to 
$D_s^{*+}\pi^0$.  We note that a similar peak is also present in 
the $M(D_s^*\pi^0)$ spectrum observed by BABAR~\cite{babar}, 
although BABAR does not claim this as evidence for a new state.  
For ease of notation, we refer to the postulated particle
as the $D_{sJ}^*(2463)^+$.

The kinematics of the $D_s^{+}\pi^0$ and $D_s^{*+}\pi^0$ decays 
are quite similar, and it is possible that they can reflect 
into one another, as noted in Ref.~\cite{babar}.  For example, taking 
$D_s^{*+}\pi^0$ events and ignoring the photon we find that essentially 
all the putative signal events form a peak in the $D_s^{+}\pi^0$ 
mass spectrum in the same location as the $D_{sJ}^*(2317)$ signal 
described in previous sections of this report. 
It is also possible that the $D_s^{+}\pi^0$ signal can pick up 
a random photon such that the $D_s^+\gamma$ combination accidentally 
falls in the $D_s^{*+}$ signal region described earlier.  In this case, 
$D_{sJ}^*(2317)^+\to D_s^+\pi^0$ decays would reflect 
into the $D_{sJ}^*(2463)^+\to D_s^{*+}\pi^0$ signal region. 
A Monte Carlo simulation of $D_{sJ}^*(2317)^+$ production and 
decay to $D_s^+\pi^0$ shows that this does happen, 
but only for approximately $9\%$ of the reconstructed events. 
By applying the same selection criteria as described in the 
preceding paragraph, except without selecting the photon from the 
$D_s^*\to D_s\gamma$ transition, we obtain $160.2 \pm 18.5$ candidates 
for the decay $D_{sJ}^*(2317)\to D_s\pi^0$.  

With this information, we can extract the number of real 
$D_{sJ}^*(2317)^+\to D_s^{+}\pi^0$ events in our data, 
denoted as $R_0$, as well as the number of real 
$D_{sJ}^*(2463)^+\to D_s^{*+}\pi^0$ events, denoted as 
$R_1$, taking into account that the real signal events 
can both feed each other.  The observed numbers of events 
are $N_0=160.2 \pm 18.5$ $D_s \pi^0$ events and $N_1 =53.3\pm 9.7$ 
$D_s^* \pi^0$ events.
The following equations relate the real to observed numbers:
\begin{equation}
  N_0 = R_0 + R_1 * f_1
\end{equation}
\begin{equation}
  N_1 = R_1 + R_0 * f_0
\end{equation}
where $R_x$ is number of real events produced times
the efficiency to observe them as signal, and $f_x$ is
the probability to feed up or feed down relative
to the reconstruction efficiency for the respective 
signal modes.  We note that these relations represent 
first-order approximations; the higher-order corrections 
are negligible in the present case.
From our simulation we measure $f_0 = 0.0910 \pm 0.0072$ 
for the probability that a reconstructed 
$D_{sJ}^*(2317)\to D_s\pi^0$ can be combined with a 
random photon so as to mimic a $D_{sJ}^*(2463)$ candidate.  
We also obtain $f_1 = 0.840 \pm 0.044$, which includes the 
probability of feed down as well as the photon finding 
efficiency.

Solving these equations we find that 
$R_0 = 124.9 \pm 22.5$ events and $R_1 = 41.9 \pm 10.7$ events, 
where the uncertainties include both statistical and systematic 
sources.  The result for $R_1$ provides strong evidence 
for the existence of a state at 2463 MeV/c$^2$.  The statistical 
significance of the signal is estimated to be in excess of 
$5\,\sigma$ by computing the probability for the combinatoric  
background plus the feed up background to fluctuate up so as to 
give the observed yield of $53.3$ events in the peak in 
Fig.~\ref{fig:deltamstar}(a).

We have also carried out a direct estimate of the background in 
Fig.~\ref{fig:deltamstar}(a) due to $D_{sJ}^*(2317)^+ \to D_s^+\pi^0$ 
plus random photon combinations, by selecting 
events in $D_s^*$ side band regions in the $D_s \gamma\pi^0$ sample. 
The $\Delta M$ distribution for this sample, shown in 
Fig.~\ref{fig:deltamstar}(b), shows only a small enhancement in the 
region of the $D_{sJ}^*(2463)$, thereby demonstrating that the 
background from $D_{sJ}^*(2317)$ decays indeed constitutes only a  
small fraction of the 53.3 events observed in the $D_{sJ}^*(2463)$ peak.
Performing a $\chi^2$ fit to the $D_s^*$-side-band-subtracted version of 
Fig.~\ref{fig:deltamstar}, we obtain consistent results 
with $R_1 = 40.8\pm 11.3$ events and a value for 
the $D_{sJ}^*(2463)-D_s^*$ mass difference of $351.6 \pm 1.7\;$MeV/c$^2$.
By comparing the values of $\chi^2$ obtained with and without the 
$D_{sJ}^*(2463)$ signal contribution, we infer that the statistical 
significance of the signal is $4.4\,\sigma$.  
From our fits to data and Monte Carlo $\Delta M$ distributions, we also infer 
a $90\,\%$ confidence level (C.L.) upper limit on the natural width 
($\Gamma$) of the $D_{sJ}^*(2463)^+$ state to be $7\;$MeV.

Based on the event yields and detection efficiencies given above, 
we can determine the production rate times branching fraction 
for the $D_{sJ}^{*}(2463)$ state to that of the $D_{sJ}^{*}(2317)$.  
We find this to be approximately $40\,\%$.

Unlike the case of the $0^+$ state, the 
$D_s\pi^+\pi^-$ decay mode is allowed by parity and angular 
momentum conservation for a state with $J^P = 1^+$.  In the model 
of Ref.~\cite{bardeeneichtenhill}, it is predicted to occur with 
a branching fraction of $19\,\%$ relative to that for $D_s^*\pi^0$. 
We have fit the $\Delta M_{D_s\pi\pi}$ spectrum plotted in 
Fig.~\ref{fig:deltampipi} for a signal corresponding 
to a transition from the $D_{sJ}^*(2463)$. Such a signal 
would peak with a $\Delta M_{D_s\pi\pi}$ near 495 MeV/c$^2$ and a Gaussian 
width of $3.4\pm 0.2\;$MeV/c$^2$ assuming a natural width of zero.  
No evidence for a signal is found, and we obtain an upper limit 
on the relative branching fraction of $8.1\,\%$ at the $90\%$ C.L.


Having obtained evidence for the $D_{sJ}^*(2463)$ state, and 
having characterized the background that it contributes in the 
$D_s\pi^0-D_s$ mass difference spectrum, we are now able to further 
address properties of the $D_{sJ}^*(2317)$ state.  
We recall that our measurement of the Gaussian width of the peak 
in Fig.~\ref{fig:manddeltam}, 
$8.4^{+1.4}_{-1.2}\;$MeV/c$^2$, 
is somewhat larger than our mass difference resolution,  
$6.4\pm 0.4\;$MeV/c$^2$.  This difference is consistent with 
predictions from Monte Carlo simulations where we include both 
$D_{sJ}^*(2463)$ and $D_{sJ}^*(2317)$ production, 
since roughly one third of the observed $D_s^+\pi^0$ events 
in the $D_{sJ}^*(2317)$ signal peak enter as a reflection from 
the $D_{sJ}^*(2463)$ state, this `background' peak having an 
expected Gaussian width of $14.9\pm 1.0\;$MeV/c$^2$.

To better determine the mass and natural width of this state, 
we carry out a binned likelihood fit of the peak in the 
$\Delta M$ spectrum in Fig.~\ref{fig:manddeltam}(b) to a 
sum of two Gaussians, one for the $D_{sJ}^*(2317)$ signal, 
and one to account for the feed down from the $D_{sJ}^*(2463)$.  
Allowing the means and widths of both Gaussians to float, 
we measure $\overline{\Delta M}$ for the $D_{sJ}^*(2317)$ to be  
$ 350.4\pm 1.2\;$MeV/c$^2$ with a width of $5.5\pm 1.3\;$MeV/c$^2$.  The 
mean and width for the feed down contribution are 
$349.2\pm 3.6\;$MeV/c$^2$ and $15.3\pm 4.1\;$MeV/c$^2$, respectively.
Both widths are consistent with predictions from Monte Carlo 
simulations in which the two states are modeled as having 
a natural width of zero.  

We have also carried out fits in which one or both of 
the widths of the Gaussians were fixed to values determined 
by the Monte Carlo.  In all cases the results were consistent 
with the results from the fit described above.  These 
two-Gaussian fits return similar values for 
$\overline{\Delta M}$ for the $D_{sJ}^*(2317)$ to the value 
given above.  
We have also tried to obtain a purer $D_{sJ}^*(2317)$ sample 
by vetoing events with photons that can be combined with the 
$D_s$ candidate to form a $D_s^*$, thereby removing some 
of the feed down background from the $D_{sJ}^*(2463)$.  
This veto marginally improves the $D_s\pi^0$ signal 
when we fit with two Gaussians, and the mass and width 
change by only a small fraction of the statistical uncertainty.
The systematic uncertainty on $\overline{\Delta M}$ receives 
contributions from uncertainties in the characterization  
of the $D_{sJ}^*(2463)$ feed down and from uncertainties in 
the modeling of the energy resolution of the calorimeter. 
Conservatively, we estimate the total systematic error to 
be 1.0 MeV/c$^2$, however further study should allow the  
size of the estimated error to be decreased.  Based on these  
studies, we limit the natural width of the $D_{sJ}^*(2317)$ 
to be $\Gamma < 7\;$ MeV at $90\%$ C.L.  


With regard to the alternate $D_{sJ}^*(2317)$ decay channels 
described earlier,  in which no signals were observed, 
we summarize the limits on the 
branching fractions relative to the $D_s^+\pi^0$ mode 
in Table~\ref{tab:limits}.  
The normalization for these limits is based on the determination 
that $78.1\pm 13.9\,\%$ 
of the observed yield of $231^{+31}_{-29}$ 
events in the peak of the $\Delta M(D_s\pi^0)$ spectrum 
in Fig.~\ref{fig:manddeltam} are 
attributable to $D_{sJ}^*(2317)\to D_s\pi^0$ decay after 
accounting for the feed down from decays of the $D_{sJ}^*(2463)$ 
state to $D_s^*\pi^0$.    
\begin{table}
  \caption{\label{tab:limits} 90\% CL upper limits on the ratio
  of branching fractions for $D_{sJ}^*(2317)$ to the the channels 
  shown relative to the $D_s^+\pi^0$ state.  Also shown are the 
  theoretical expectations from Ref.~\cite{bardeeneichtenhill}, 
  under the assumption that the $D_{sJ}^*(2317)$ is the 
  lowest-lying $0^+$ $c\overline{s}$ meson.}
  \begin{ruledtabular}
  \begin{tabular}{lrccc}
  Final State & Yield & Efficiency & Limit ($90\%$ CL) & Prediction \\
  \hline
  $D_s^+ \pi^0 $           & $180\pm 46$   & $(13.1 \pm 0.7)\,\%$ &  --- 
                           & \\
  $D_s^+ \gamma$           & $-22\pm 13$   & $(18.4 \pm 0.9)\,\%$ & $< 0.054 $ 
                           & 0 \\
  $D_s^{*+} \gamma$        & $-2.0\pm 4.1$ & $( 5.3 \pm 0.4)\,\%$ & $< 0.078 $ 
                           & 0.08 \\
  $D_s^+ \pi^+\pi^-$       & $ 1.6\pm 2.6$ & $(19.6 \pm 0.7)\,\%$ & $< 0.020 $ 
                           & 0 \\
  \end{tabular}
  \end{ruledtabular}
\end{table}
The event yields are obtained by fitting the mass difference distributions 
to a Gaussian with mean fixed to the result from the $D_s^+\pi^0$ channel 
and width specified by the resolution determined from 
the simulation of the corresponding decay mode.  
Uncertainties are dominated by the statistical error on the
unseen yields and limits on the relative rates
are calculated assuming a Gaussian distribution
with negative values not allowed.

 

  In summary, data from the CLEO~II detector provides confirming 
evidence for the existence of a new narrow resonance decaying to 
$D_s^+\pi^0$, with a mass near 2.32 GeV/c$^2$.  This state is consistent 
with being the $0^+$ member of the lowest-lying $P$-wave $c\overline{s}$ 
multiplet.  
We have not observed other decay modes of this state, as summarized in
Table~\ref{tab:limits}.  We have measured the mass splitting of this 
state with respect to the $D_s$ meson to be 
$ 350.4\pm 1.2\;\mbox{\rm [stat.]}\pm 1.0\;\mbox{\rm [syst.]}\;$MeV/c$^2$, 
and we find its natural width to be $\Gamma < 7\;$MeV at $90\%$ C.L.

We have observed a second narrow state with a 
mass near 2.46 GeV/c$^2$, decaying to $D_s^{*+}\pi^0$.  The measured 
properties of this state are consistent with its interpretation as 
the $1^+$ partner of the $0^+$ state in the spin multiplet with 
light quark angular momentum of $j=1/2$.  
We have measured the mass splitting of this 
state with respect to the $D_s^*$ meson to be 
$ 351.6\pm 1.7\;\mbox{\rm [stat.]}\pm 1.0\;\mbox{\rm [syst.]}\;$MeV/c$^2$.
The natural width of this state is also found to be $\Gamma < 7\;$MeV 
at $90\%$ C.L.  Since the $D_{sJ}^{*}(2463)$ mass lies above the 
kinematic threshold for decay to $DK$ (but not for $D^*K$), the narrow 
width suggests that this decay does not occur. This is additional 
evidence for the compatibility of the $D_{sJ}^{*}(2463)$ with the 
$J^P = 1^+$ hypothesis.  

In the model of Bardeen, Eichten and Hill, a $J^P = 1^+$ state is 
predicted with the same mass splitting $\Delta M$ with the $1^-$ state 
as that between the $0^+$ and $0^-$ states.  Taking the difference 
between the two mean mass differences reported above, 
we obtain $\delta(\Delta M) = (351.6\pm 1.7) - (350.4\pm 1.2) = 
1.2 \pm 2.1\;$MeV/c$^2$ for the difference between the
$1^+ - 1^-$ and $0^+ - 0^-$ mass splittings, 
where the dominant uncertainty is due to statistics.  
Thus our observations are consistent with these predictions.


We gratefully acknowledge the effort of the CESR staff 
in providing us with excellent luminosity and running conditions.
We thank W.\ Bardeen, E.\ Eichten, S.\ Godfrey, C.\ Hill  
and J.\ Rosner for useful discussions.
This work was supported by the National Science Foundation,
the U.S. Department of Energy, the Research Corporation,
and the Texas Advanced Research Program.


\end{document}